\documentclass[reprint, amsmath, amssymb, aps,
 pra, floatfix, superscriptaddress]{revtex4-2}
\usepackage{siunitx}
\usepackage{graphicx}
\usepackage{dcolumn}
\usepackage{bm}
\usepackage{sidecap}
\usepackage{booktabs}

\usepackage{multirow}
\usepackage{makecell}
\usepackage{threeparttable}
\begin{document}

\preprint{APS/123-QED}

\title{Machine Learning–Accelerated SSNEB for Efficient Minimum Energy Pathway Calculations}

\author{Yu Zhang}
 \affiliation{Department of Physics, University of Florida, Gainesville, FL 32622, USA.}
 
\author{Guanzhi Li}%
\affiliation{%
  Stanford Institute for Materials and Energy Sciences, SLAC National Accelerator
Laboratory, Menlo Park, CA 94025, USA.
}%

\author{Minkyung Han}
\affiliation{%
  Stanford Institute for Materials and Energy Sciences, SLAC National Accelerator
Laboratory, Menlo Park, CA 94025, USA.
}%
\affiliation{
Department of Earth and
Planetary Sciences, Stanford University, Stanford, CA 94305, USA.
}%

\author{Sean Gasiorowski}
\affiliation{
SLAC National Laboratory, Menlo Park, CA 94025, USA.
}%

\author{Daniel Ratner}
\affiliation{
SLAC National Laboratory, Menlo Park, CA 94025, USA.
}%

\author{Chunjing Jia}
\email{cjia1@ufl.edu}
\affiliation{Department of
Physics, University of Florida, Gainesville, FL 32622, USA.}

\author{Yu Lin}
\email{lyforest@slac.stanford.edu}
\affiliation{%
  Stanford Institute for Materials and Energy Sciences, SLAC National Accelerator
Laboratory, Menlo Park, CA 94025, USA.
}%

\date{\today}

\begin{abstract}
  Metastable states and their minimum energy pathways (MEPs) are central to understanding transformations and phase stability in complex materials, yet mapping transition pathways between competing states remains computationally demanding and experimentally challenging. Here, we introduce a hybrid solid-state nudged elastic band (SSNEB) framework that integrates two pretrained machine learning models, EquiformerV2 (eqV2)~\cite{barroso_omat24} and the equivariant Smooth Energy Network (eSEN) ~\cite{fu_esen}, with DFT for energy, force, and stress evaluations. Applied to three solid-state systems, CsPbI{\textsubscript{3}}, GaN, and TiO{\textsubscript{2}}, our framework achieves up to a 7-fold speedup while converging to the same pathways predicted by first-principles calculations. Moreover, the hybrid SSNEB framework enables systematic benchmarking of existing ML models, providing both efficiency and reliability for predicting MEPs across various materials.

\end{abstract}

\maketitle


\section{\label{sec:level1}Introduction}

Understanding the minimum energy pathways (MEPs) connecting states and structures is a fundamental problem in materials science, physics, and chemistry \cite{PhysRevB.96.165305, Schaaf2023, Kim2024}. These pathways provide insights into transition mechanisms, energy barriers, kinetic properties, and metastable intermediate states that can be stabilized under external conditions such as high pressure and temperature. Metastable phases often exhibit properties superior to those of their thermodynamically stable counterparts, creating opportunities for the design of advanced materials and the discovery of emergent states and functionalities for applications ranging from photovoltaics \cite{PSC} and superconductors \cite{Superconductor} to superhard and lightweight structural materials \cite{metastability1,metastability2}. Efficiently exploring this vast landscape of stable and metastable states, particularly across a wide pressure–temperature (\textit{P-T}) phase space, remains a central challenge in materials discovery. 


The nudged elastic band (NEB) method is a widely used tool for probing MEPs by connecting the initial and final states through a chain of intermediate images and optimizing the pathway under artificial spring constraints. Conventional NEB is formulated primarily for molecular or finite systems, making it less suitable for periodic solids, where lattice degrees of freedom and stress tensors must also be considered. To address this limitation, \citet{G-SSNEB} generalized the definition of geometric distances and forces in NEB to explicitly include lattice vectors and stress tensors (i.e., forces acting on the lattice vectors), and introduced the solid-state nudged elastic band (SSNEB) method.


\begin{figure}[hbtp]
  \centering
  \includegraphics[width=\columnwidth]{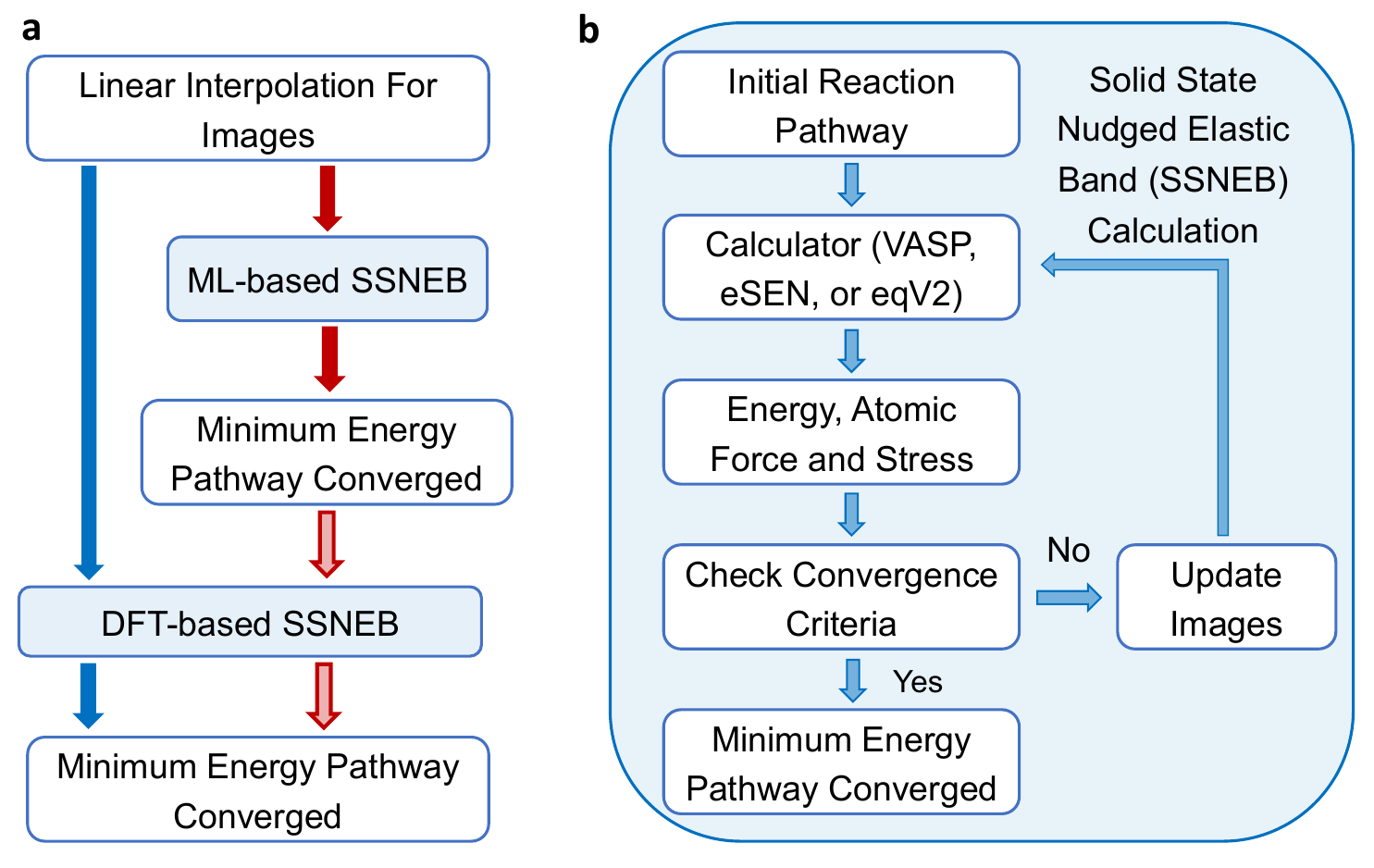}
  \caption{Flowchart of our ML-based SSNEB approach. {\bf a}, blue arrows indicate the traditional approach, where intermediate images are generated by linear interpolation, followed by DFT calculations. Red arrows represent our ML-based SSNEB, where MEP is first obtained using model-only calculations (solid arrow), and then refined by restarting DFT-based SSNEB from model converged images (hollow arrow). {\bf b}, Algorithm of the SSNEB method.}
  \label{flowchart}
\end{figure}

Despite its advantages for modeling MEPs in periodic systems \cite{PhysRevB.100.224101, doi:10.1126/sciadv.ado9593,osti_1978706}, SSNEB is computationally demanding because each image update requires accurate evaluations of energies, forces, and stresses, which are typically obtained from first-principles density functional theory (DFT) calculations. Recent advances in machine learning (ML) interatomic models \cite{barroso_omat24, fu_esen, Mace}, trained on large datasets of inorganic materials and adapted for downstream atomistic tasks, offer a promising route to overcome this computational bottleneck. However, their accuracy and reliability require careful evaluations. ML surrogates have been widely employed to accelerate conventional NEB calculations, for example, through Gaussian process models and neural network potentials that reduce the number of expensive DFT evaluations along the pathway \cite{Garrido2019,fantasia2024development}. Nevertheless, the explicit integration of ML models into the SSNEB framework remains, to our knowledge, unexplored.

Here, we present an integrated ML-DFT framework tailored for SSNEB calculations, where pretrained machine learning models are used as DFT surrogates for energy, force, and stress evaluations. Through tests on transition pathways for several material systems, including  CsPbI{\textsubscript{3}}, GaN, and TiO{\textsubscript{2}}, we demonstrate the validity of our framework and gain insight into the strengths and limitations of existing ML models.








\section{Methods} 
\label{gen_inst}

Figure \ref{flowchart}a illustrates the overall flow of our integrated ML-based SSNEB approach, where Fig. \ref{flowchart}b shows the algorithmic flow of SSNEB. Our ML-based SSNEB framework modifies the conventional SSNEB workflow by integrating ML models into the pathway search. In the traditional method, the intermediate images are generated by linear interpolation and evaluated solely with DFT, which is computationally expensive. In contrast, our approach first employs ML models to compute the MEP based exclusively on model evaluations. Once this approximate pathway is obtained, the calculation is refined by restarting SSNEB with DFT, using the ML-converged images as the initial pathway. This hybrid workflow reduces the overall computational cost while preserving the accuracy of DFT-based pathways.
 
In this work, we employ two equivariant neural network architectures, EquiformerV2 (eqV2)~\cite{barroso_omat24} and the equivariant Smooth Energy Network (eSEN) ~\cite{fu_esen}, as ML surrogates for DFT, serving as calculators for energy, force, and stress calculations. For eqV2, we used the 153 million-parameter variant (eqV2-L-OMat), pretrained on the Open Materials 2024 (OMat24) dataset~\cite{barroso_omat24}. For eSEN, we used the 30 million-parameter model (eSEN-30M-OMat), also pretrained on the OMat24 dataset. All model predictions were performed using a random seed of 42.

For SSNEB calculations (Fig. \ref{flowchart}b), each image was evaluated for energies, atomic forces, and stresses using the chosen calculator, enabling either DFT- or ML-based workflow. A combination of forces on the atoms, including spring forces along the pathway and forces related to the stress and lattice degrees of freedom, was considered. The total force on an atom is given by:
$F^{\text{NEB}}=F^{\text{spring}\parallel}+F^{\text{ (scaled stress+atomic forces)}\perp}$.
The SSNEB calculations were considered to have converged when the total force on any atom in any image was less than \SI{0.01}{\electronvolt/\text{Å}}. For DFT-based calculations, all electronic structure computations were performed using the Vienna Ab-initio Simulation Package (VASP)~\cite{VASP}. For pressure-dependent SSNEB calculations, the enthalpy was evaluated using a modified formulation that explicitly incorporates the \textit{PV} contribution. Additional details can be found in the Appendix.

\section{Results} 
\label{headings}

\begin{figure*}[htbp!]
  \centering
  \includegraphics[width=1.0\linewidth]{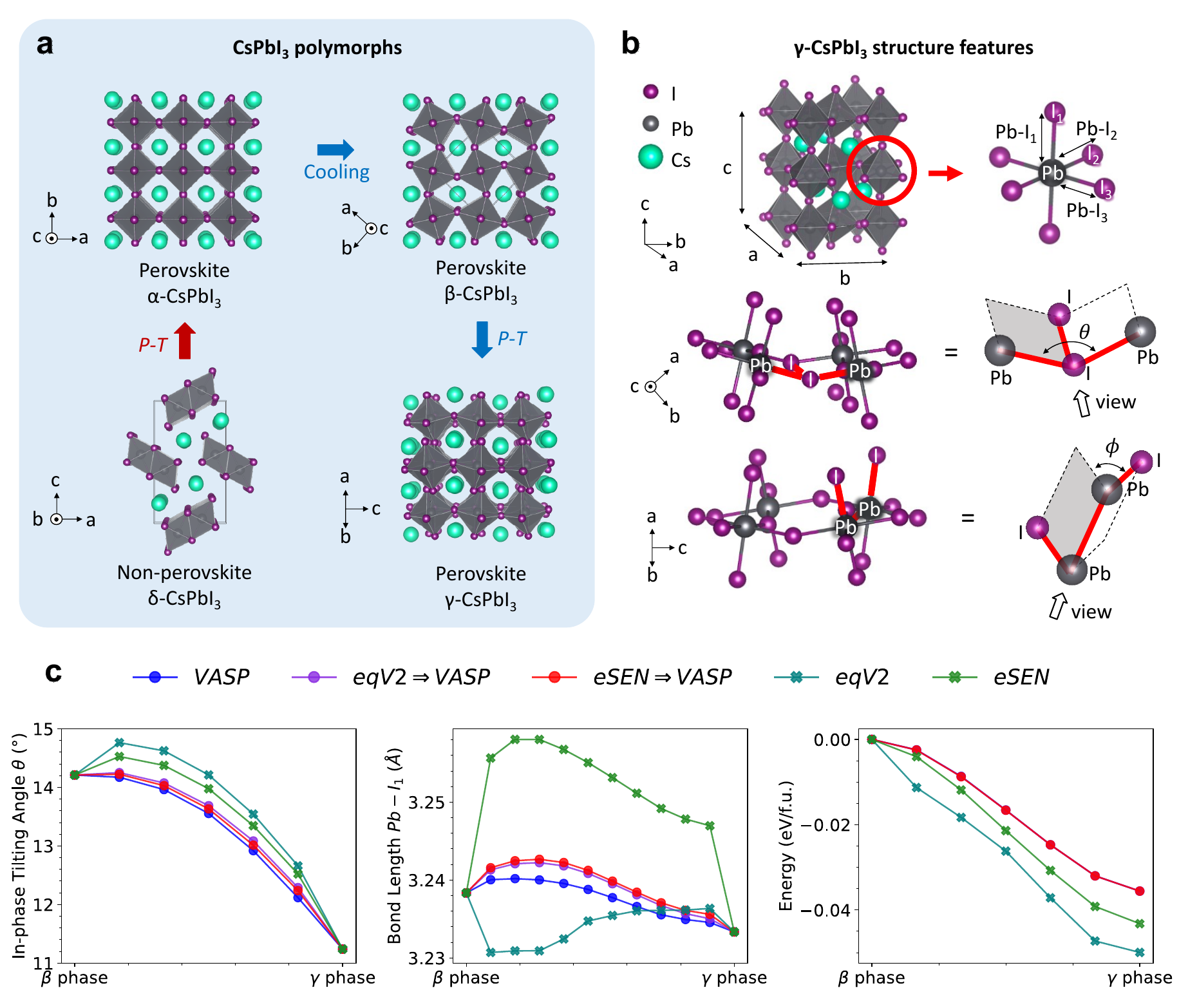}
  \caption{Crystal structures and SSNEB calculations for CsPbI{\textsubscript{3}}. {\bf a}, Crystal structures of the $\alpha$, $\beta$, $\gamma$, and $\delta$ phases, and their phase transitions under varying pressure (\textit{P}) and temperature (\textit{T}). {\bf b}, Key structural features of the $\gamma$ phase, including Pb–I bond lengths, the in-phase octahedral tilting angle ($\theta$), and the out-of-phase tilting angle ($\phi$). The figure is reproduced from Han \emph{et al.}\cite{han2024machine}. {\bf c}, SSNEB calculations using five approaches: VASP-based, eqV2 with subsequent VASP restart,  eSEN with subsequent VASP restart, eqV2-based, and eSEN-based. Results are shown for the in-phase tilting angle $\theta$, Pb–I$_1$ bond length, and MEPs.}
  \label{CsPbI3_plots}
\end{figure*}

We validated the framework using three selected test systems, a halide perovskite cesium lead iodide (CsPbI{\textsubscript{3}}), gallium nitride (GaN), and titanium dioxide (TiO{\textsubscript{2}}). These systems were chosen for their scientific interest, the diversity of their stable and metastable phases, the availability of prior knowledge, and the feasibility of experimental validation, therefore enabling an iterative and reliable evaluation process. 

\subsection{CsPbI{\textsubscript{3}} results}


Halide perovskites represent an extensive family of materials that have received extraordinary research attention due to their remarkable structural, optical, and electronic properties, enabling applications in a wide range of technologies such as high-efficiency solar cells\cite{doi:10.1126/science.aan2301} and photodetectors\cite{Ahmadi2017}. Structurally, 3D halide perovskites have the general chemical formula ABX\textsubscript{3} (X = Cl, Br, I) and adopt a CaTiO\textsubscript{3} structure, where the anionic network of corner-sharing B-X octahedra is charge balanced by small organic or inorganic A-site cations~\cite{kojima2009organometal}. Their soft lattices, characterized by low bulk moduli, result in dramatic structural and property changes under external stimuli such as pressure~\cite{jaffe2017halide}.

CsPbI\textsubscript{3}, an all-inorganic halide perovskite, is particularly attractive because its 3D perovskite phases exhibit optimal band gaps suitable for photovoltaic applications along with chemical stability against heat and humidity. However, the small size of the Cs\textsuperscript{+} ion induces phase instability, where the three perovskite “black” phases with corner-sharing octahedra - cubic $\alpha$-phase, tetragonal $\beta$-phase, and orthorhombic $\gamma$-phase - spontaneously transform to the non-perovskite “yellow” $\delta$-phase with edge-sharing octahedra at room temperature, losing the functionality (Fig. \ref{CsPbI3_plots}a)~\cite{stoumpos2013semiconducting}.

Prior experimental work presented a high-pressure strategy to engineer octahedral tilting and manipulate the phase (meta) stability of CsPbI\textsubscript{3} for the synthesis and recovery of the metastable perovskite $\delta$-phase with enhanced functionality to ambient conditions~\cite{ke2021preserving}. Complementary ML work further validated the predictive power of various models, especially graph neural networks, in capturing key physical properties such as bandgap and enthalpy of CsPbI\textsubscript{3} at high pressure, establishing it as a benchmark system for ML-driven materials discovery~\cite{han2024machine}.

\begin{table}[htbp!]
\centering
\caption{Comparison of the number of VASP iterations required for SSNEB convergence before and after applying ML models for the CsPbI\textsubscript{3} system.}
\label{CsPbI3}
\begin{tabular}{cccc}
\toprule
CsPbI$_3$ &  & \multicolumn{2}{c}{Convergence Iterations} \\
\midrule
Transition & Pretrained model & $N_{\mathrm{img}}=5$ & $N_{\mathrm{img}}=10$ \\
\midrule
$\alpha \rightarrow \beta$ & eSEN & $56 \rightarrow 25$  & $59 \rightarrow 28$  \\
$\alpha \rightarrow \beta$ & eqV2 & $56 \rightarrow 25$  & $59 \rightarrow 28$  \\
$\beta \rightarrow \gamma$ & eSEN & $125 \rightarrow 73$ & $235 \rightarrow 81$ \\
$\beta \rightarrow \gamma$ & eqV2 & $125 \rightarrow 86$ & $235 \rightarrow 132$\\
$\alpha \rightarrow \gamma$ & eSEN & $186 \rightarrow 46$ & $370 \rightarrow 55$ \\
$\alpha \rightarrow \gamma$ & eqV2 & $186 \rightarrow 58$ & $370 \rightarrow 81$ \\
\bottomrule
\end{tabular}
\begin{tablenotes}
    \item $N_{\mathrm{img}}$ represents the number of intermediate states.
    \item For each transition, linearly interpolated images were replaced with the ML-converged pathway. The reported values are given in the form $X \rightarrow Y$, where $X$ denotes the number of iterations required for convergence using DFT only, and $Y$ denotes the number of DFT steps required after restarting from the ML-converged pathway. A two to threefold reduction is observed in the last two columns.
\end{tablenotes}
\end{table}

In this work, we accelerate the search for optimal phase transition pathways by combining pretrained ML models on large material databases with the SSNEB method. CsPbI\textsubscript{3} is particularly well suited for this study because of its complex energy landscape with multiple competing phases at local minima, which complicates transition path predictions. Figure \ref{CsPbI3_plots}c presents our ML-based SSNEB benchmarks for the ambient-pressure $\beta \rightarrow \gamma$ phase transition of CsPbI\textsubscript{3}. Additional results for $\alpha \rightarrow \beta$ and $\alpha \rightarrow \gamma$ could be found in the supplementary material. The VASP-converged path serves as the high-resolution reference or ground truth. The ML-converged pathways differ from VASP results in terms of the MEP and structural attributes (\textit{i.e.}, octahedral tilting angles, bond lengths). Then, we restart DFT-based SSNEB from ML-converged paths, both (eqV2 and eSEN) converged to VASP results with fewer SSNEB iterations, as shown in Table \ref{CsPbI3}. On average, the integration of eSEN and eqV2 models reduces the number of required DFT calculations to 30\% and 40\%, respectively, indicating that the eSEN model performs better. Therefore, we focus on the eSEN model for the following tests. Additionally, as the number of intermediate states, $N_{\mathrm{img}}$, increases from 5 to 10, the ratio of DFT iterations required for convergence by the hybrid SSNEB method relative to the conventional SSNEB method remains comparable or decreases, with most cases showing a further reduction, indicating enhanced efficiency of the hybrid approach (Table \ref{CsPbI3}).







\begin{figure*}[htbp!]
  \centering
  \includegraphics[width=1.0\linewidth]{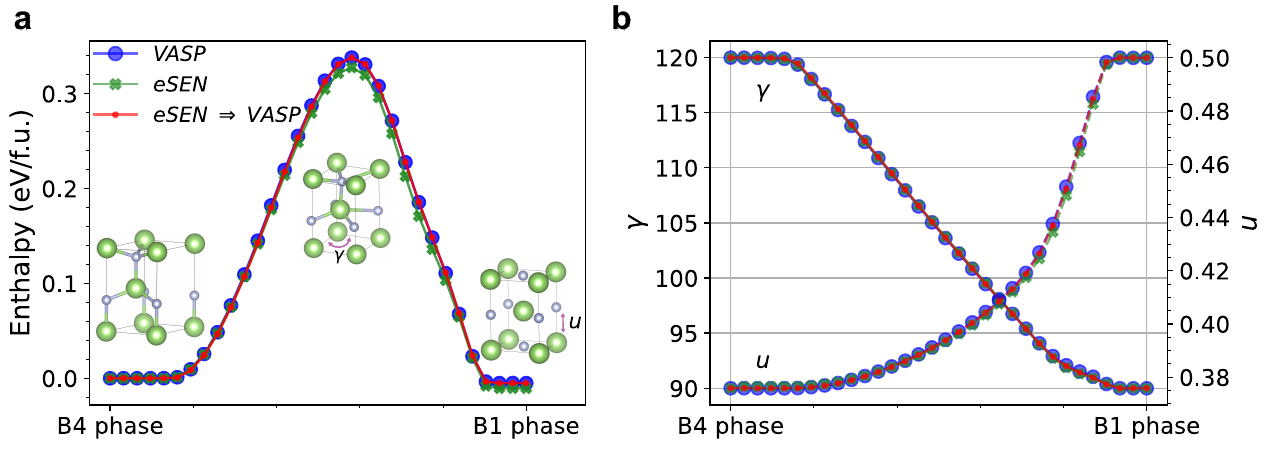}
  \caption{{\bf a}, The transition pathways calculated using the SSNEB method, showing an enthalpy barrier of 0.34 eV/f.u. under the transition pressure $P_t = 45.7$ GPa. Insets show the crystal structures corresponding to the initial state, saddle point, and final state. Green and gray spheres represent Ga and N atoms, respectively. {\bf b}, Evolution of the structural parameters $\gamma$ and $u$ along the pathways.}
  \label{GaN}
\end{figure*}

\subsection{GaN results}

Gallium nitride (GaN) is a widely studied wide-bandgap semiconductor. Its structural evolution under high pressure, particularly the pressure-induced wurtzite-to-rocksalt transition (B4-to-B1 transition), has been extensively investigated in recent years due to the complex mechanism involved in this phase transition \cite{VCNEB, GaN}. During the transition, two possible intermediate phases, tetragonal and hexagonal, may emerge. These phases can be characterized by changes in the structural parameters $u$ (from $0.377$ to $0.5$) and $\gamma$ (from $120^\circ$ to $90^\circ$). Here, $u$ describes the ratio between the Ga-N bond length and the lattice constant $c$, while $\gamma$ denotes the angle between lattice vectors ${\vec{\bf a}}$ and ${\vec{\bf b}}$.

Both $\gamma$ and $u$ exhibit distinct evolution behaviors along the transition pathway from the B4 to the B1 phase. After the plateau region, $\gamma$ decreases rapidly from $120^\circ$ toward the transition state, followed by a more gradual decrease as the structure approaches the B1 phase. In contrast, $u$ changes only slightly in the early stages but increases more rapidly near the B1 phase. At the transition state, $\gamma$ and $u$ are 100.84$^\circ$ and 0.40, respectively. Compared with the values for the B4 phase ($u = 0.38$ and $\gamma = 120^\circ$), the change in $\gamma$ is substantially larger than that in $u$. These results suggest that lattice-angle distortion precedes bond-length evolution, supporting a tetragonal-like transition pathway.

Our ML-assisted SSNEB workflow achieved significant performance improvement. First, the eSEN-converged pathway reproduces a comparable enthalpy barrier to that obtained from pure VASP calculations, with values of 0.33 and 0.34 eV/formula, respectively. Furthermore, after restarting VASP from the eSEN-converged pathway, the number of required SCF calculations was reduced to only 13\% of that required for the fully VASP-driven calculation, while converging to an equivalent result.

\begin{table}[htbp!]
\centering
\caption{Comparison of the number of VASP iterations required for SSNEB convergence before and after applying ML models for GaN and TiO\textsubscript{2} systems.}

\begin{tabular}{cccc}
\toprule
 &  & \multicolumn{2}{c}{Convergence Iterations} \\
\midrule
Transition & Pretrained model  &  $N_{\mathrm{img}}=5$ & $N_{\mathrm{img}}=30$\\
\midrule
GaN B4-B1 & eSEN & $89 \rightarrow 12$  & $361 \rightarrow 47$\\
TiO\textsubscript{2}-B to anatase & eSEN & $80 \rightarrow 58$ & \\
\bottomrule
\end{tabular}

\end{table}

\subsection{TiO\textsubscript{2} results} 

\begin{figure}[htbp!]
  \centering
  \includegraphics[width=1.0\linewidth]{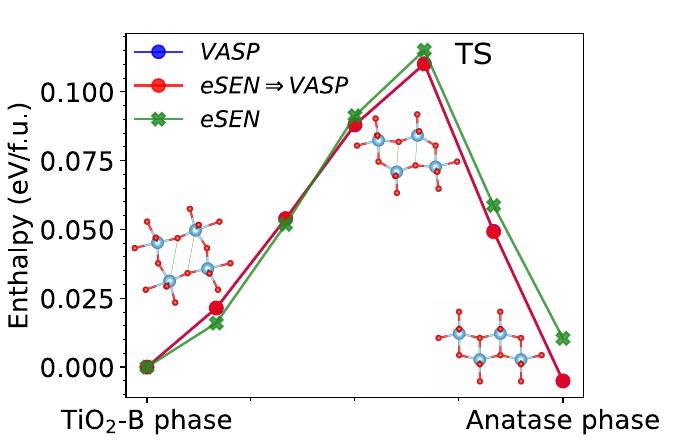}
  \caption{Calculated MEPs from TiO\textsubscript{2}-B to anatase phase using three approaches: DFT-based SSNEB, eSEN with subsequent VASP restart, and eSEN-based SSNEB. VASP-only and eSEN with subsequent VASP converge to the identical MEP. Insets show the crystal structures of the TiO\textsubscript{2}-B phase, the transition state at the energy barrier, and the anatase phase. Red and blue spheres represent O and Ti atoms, respectively.}
\end{figure}

As another example, we calculate the MEP for TiO$_2$, a widely studied transition-metal oxide known for its rich polymorphism and application as a photocatalyst, including phases such as TiO$_2$-B and anatase \cite{TiO2}. Figure 4 presents the MEPs obtained using three approaches: DFT-based SSNEB, the eSEN method with subsequent VASP restart, and eSEN-based SSNEB. All three methods capture the key features of the TiO$_2$-B to anatase transformation, with the transition state at the saddle point of the energy barrier clearly resolved. The VASP-based approach provides a high-resolution enthalpy barrier of 0.88 eV using the climbing-image SSNEB, but it is also the most computationally demanding. The eSEN-based pathway yields a slightly higher barrier of 0.92 eV, yet still captures the qualitative feature of the transition at a dramatically reduced cost, as the inference time is much shorter than DFT computational time (see Table \ref{Computational_Time} for more details). In contrast, the eSEN + VASP restart approach reproduces an energy barrier identical to the VASP-only result (0.88 eV) while reducing the number of DFT calculations by a factor of 0.73. These comparisons demonstrate that incorporating eSEN into the workflow offers an efficient strategy for exploring complex phase transitions without compromising accuracy.




\section{Discussion and outlook} 
\label{others}
Our results establish that pretrained eqV2- and eSEN-based models enable robust convergence of SSNEB calculations, allowing reliable identification of both transition pathways and associated energy barriers. Between the two, eSEN consistently demonstrates superior performance, offering greater stability and accuracy in mapping the pathway. Importantly, when restarted with VASP, the SSNEB calculations produce results comparable to high-resolution VASP-only simulations, while achieving a 2–7-fold improvement in computational efficiency. This efficiency gain is especially striking given that the models were applied without any fine-tuning or transfer learning tailored to the specific materials under investigation. These findings highlight the promise of ML-based SSNEB frameworks as scalable tools for accelerating transition-state searches in complex materials. 

Looking ahead, fine-tuning with DFT in active-learning loops~\cite{NN-BAX} or with {\it ab initio} molecular dynamics offers a natural path to further improve efficiency, particularly for material-specific systems. A key challenge will be optimizing the relative weights in the model to balance the accuracy of the energies, forces, and stresses, thus improving the quality of the total forces of the SSNEB. With the rapid advances in ML, we also anticipate that newer pretrained models will surpass current ones and, when integrated, further enhance the ML-based SSNEB framework. By advancing along these directions, this framework has the potential to transform how transition pathways and metastable states are resolved across functional materials with diverse chemistries and polymorphic phases stabilized under extreme conditions, ultimately guiding efficient synthesis and materials discovery.

\section*{Acknowledgments}
This work was supported by the U.S. Department of Energy, Office of Science, Basic Energy Sciences, Division of Materials Sciences and Engineering, under contract No. DE-AC02-76SF00515. Y.Z.'s summer research at SLAC, G.L., and S.G. were supported by the U.S. Department of Energy, Laboratory Directed Research and Development program at SLAC National Accelerator Laboratory, under contract DE-AC02-76SF00515. Y.Z.'s research at UF and C.J. acknowledge the support by the Center for Molecular Magnetic Quantum Materials, an Energy Frontier Research Center funded by the U.S. Department of Energy, Office of Science, Basic Energy Sciences under
Award no. DE-SC0019330. 

\nocite{*}

\bibliography{main}

\appendix

\section{Technical Appendices and Supplementary Material}

\subsection{Structure preparation}

The CsPbI\textsubscript{3} perovskite crystal structures were relaxed from experimentally measured structures \cite {doi:10.1021/acsnano.8b00267}. The GaN structures were modified and relaxed from data in Materials Project. The TiO\textsubscript{2} structures were adapted from previously reported structures in the literature \cite{VCNEB}.

\begin{table}[htbp]
\centering
\begin{threeparttable}
\caption{Computational time per structure for CsPbI$_3$ $\beta$ to $\gamma$ transition.}
\begin{tabular*}{\linewidth}{@{\extracolsep{\fill}} ccc}
\toprule
 & VASP &  eSEN\\
\midrule
32 CPUs  & 269.98s  & 0.53s \\
Single L4 GPU  & 148.85s  &  0.39s\\
Single B200 GPU &  191.49s  & 0.28s\\
\bottomrule
\end{tabular*}
    \begin{tablenotes}
        \small
        \item VASP computational time highly depends on the size and symmetry of the cell, while the inference time based on the pretrained model mostly depends on the computational resources.\\ 
    \end{tablenotes}
\label{Computational_Time}
\end{threeparttable}
\end{table}

\subsection{DFT calculations}
The DFT calculations were performed using the Vienna Ab-initio Simulation Package (VASP)~\cite{VASP}, with the projector augmented wave (PAW)~\cite{PAW} method and the Perdew-Burke-Ernzerhof (PBE)~\cite{PBE} exchange-correlation functional. For calculations of CsPbI\textsubscript{3}, a plane-wave energy cutoff of \qty{600}{\electronvolt} was used, and the convergence threshold for total energy was set to \qty{1e-8}{\electronvolt}. For the $\alpha$-to-$\beta$ phase transition, a supercell containing two formula units and a 4 $\times$ 4 $\times$ 6 Monkhorst-Pack~\cite{Monkhorst-Pack} k-point mesh was used. Due to the reduced symmetry of the gamma phase, a larger supercell with four formula units and a 4 $\times$ 4 $\times$ 3 Monkhorst-Pack k-point mesh was used for the $\alpha$-to-$\gamma$ and $\beta$-to-$\gamma$ transitions. For calculations of GaN, a plane-wave energy cutoff of 600 eV was used, and the convergence threshold for total energy was \qty{1e-5}{\electronvolt}. A cell containing 4 atoms and an 8 $\times$ 8 $\times$ 6 Monkhorst-Pack k-point mesh were used in the calculations. The external pressure is 45.7 GPa for the transformation. For calculations of TiO\textsubscript{2}, a plane-wave energy cutoff of 500 eV was used, and the convergence threshold for total energy was \qty{1e-6}{\electronvolt}. A supercell containing 24 atoms and a 4 $\times$ 4 $\times$ 4 Monkhorst-Pack k-point mesh were used in the calculations. The external pressure is 0.05 GPa.

\subsection{Enthalpy correction for SSNEB}
In the original G-SSNEB formulation \cite{G-SSNEB}, the enthalpy is expressed as:
\begin{equation}
    H = E + V_0 \sum_{i,j=1,2,3} P_{ij}\varepsilon_{ij}
\end{equation}
where $E$ is the internal energy, $P$ is external pressure, and $\varepsilon$ is strain defined as
$\varepsilon=\bf{h}^{-1}\cdot(\bf{h}^{def} - \bf h)$. For calculations performed under external pressure, we instead evaluate the enthalpy using the standard thermodynamic definition:
\begin{equation}
    H = E + PV
\end{equation}
where $V$ is the volume. This is also consistent with the enthalpy definition used in VASP.

\subsection{Dataset structure comparison}
\begin{table*}[htbp]
\centering
\caption{Comparison of structure and formula matches between our structures and the dataset.}
\begin{tabular}{lccccccc}
\toprule
Structure & CsPbI$_3$ $\alpha$ & CsPbI$_3$ $\beta$ & CsPbI$_3$ $\gamma$ & GaN B4 & GaN B1 & TiO\textsubscript{2}-B & TiO\textsubscript{2} anatase \\
\midrule
Formula match   & 20 & 20 & 20 & 565 & 565 &  408 & 408\\
Structure match & 0  & 0  & 10 & 0   & 10   &  4 &  49\\
\bottomrule
\label{StructureMatch}
\end{tabular}
\end{table*}

To determine whether the systems studied in this work were already represented in the dataset, we downloaded the OMat24 training split. We first identified candidate matches by comparing the reduced formulas of our compounds with those in the dataset. Then, for materials with matching formulas, we further evaluated the structural similarity using the \texttt{StructureMatcher} function from the \texttt{pymatgen} library. The lattice tolerance, site tolerance, and angle tolerance were set to 0.3, 0.5, and 10, respectively. The results are shown in Table \ref{StructureMatch}.



\begin{figure*}[htbp]
  \centering
  \includegraphics[width=1.0\linewidth]{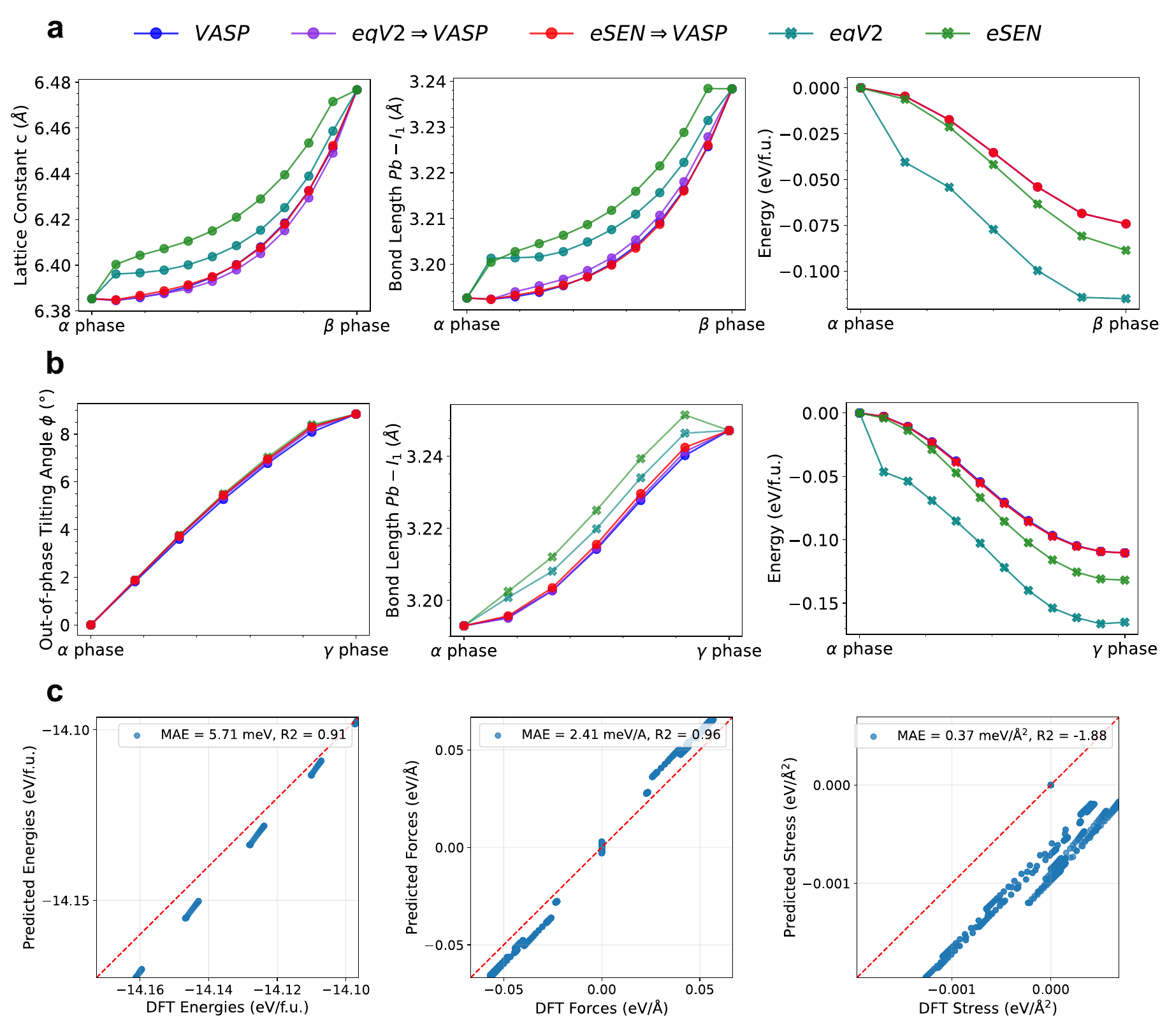}
  \caption{{\bf a}, The five CsPbI\textsubscript{3} $\alpha$ to $\beta$ transition pathways obtained from different SSNEB approaches are illustrated. The lattice constant c, Pb–I$_1$ bond length, and MEPs are presented here. The largest deviations occur in the first and last intermediate structures. {\bf b}, Structure features and MEPs for the $\alpha$ to $\gamma$ transition. {\bf c}, To directly evaluate the performance of the pretrained eSEN model, we present parity plots for energy, force, and stress. In these plots, the x-axis represents the energies, forces, and stresses calculated by VASP for the $\alpha$ to $\beta$ transition (N$_{img}$ = 5), while the y-axis shows the corresponding predictions from the eSEN model. Large deviations are observed in the stress predictions, indicating the need for further investigation.}
  \label{CsPbI3_more_features}
\end{figure*}

\end{document}